\documentclass[aps,prb,10pt,twocolumn,amsmath,amssymb,citeautoscript,
               superscriptaddress, longbibliography,noeprint,nonote]{revtex4-1}
\usepackage[utf8]{inputenc}
\usepackage{graphicx}
\usepackage{dcolumn}
\usepackage{array}
\usepackage{amssymb}
\usepackage{amsmath}
\usepackage{float}
\usepackage{multirow}
\usepackage{braket}
\usepackage[normalem]{ulem}
\usepackage{bm}
\renewcommand{\vec}[1]{\mathbf{#1}}
\usepackage[hang]{footmisc}
\usepackage{hyperref}
\usepackage{orcidlink}
\hypersetup{pdfnewwindow=true, colorlinks=true, linkcolor=blue, anchorcolor=blue, citecolor=blue, filecolor=blue, menucolor=blue, urlcolor=blue}

\newcommand{\sm}[1]{the SM\cite{[See Supplemental Material at ][{ #1}]SM}}

\def\Fclmbelec{\epsilon}               %
\def\Fclmb{\bar{\Fclmbelec}}    %
\def\Fcboaelec{U}               %
\def\Fcboa{\bar{\Fcboaelec}}    %
\def\Fefieldelec{\mathcal{F}}             %
\def\Fefield{\bar{\Fefieldelec}}%
\def\elecG{\psi}                       %
\def\elecM{\bar{\elecG}}            %
\def\elecC{\elecM}                  %
\def\elecQ{\elecM_{\Fcboaelec}}     %
\def\elecE{\elecM_{\Fefieldelec}}   %
\def\dipole{\mu}                       %
\def\dipoleop{\hat{\dipole}}         %
\def\dipoleelec{\dipole}             %
\def\dipoleM{\bar{\dipole}}          %
\def\dipoleQ{\dipoleM_{\Fcboaelec}}  %
\def\dipoleE{\dipoleM_{\Fefieldelec}}%

\begin{document}
\title{Cavity-Born Oppenheimer Approximation for Molecules and Materials via Electric Field Response}
\author{John Bonini \orcidlink{0000-0003-2865-3807}}
\affiliation{Material Measurement Laboratory, National Institute of Standards and Technology,
100 Bureau Dr, Gaithersburg, MD, USA.}
\affiliation{Center for Computational Quantum Physics, Flatiron Institute, 162 5th Avenue, New York, New York 10010, USA}
\author{Iman Ahmadabadi \orcidlink{0000-0001-9703-2241}}
\affiliation{Center for Computational Quantum Physics, Flatiron Institute, 162 5th Avenue, New York, New York 10010,
	USA}
\affiliation{Joint Quantum Institute, NIST and University of Maryland, College Park, Maryland 20742, USA}
\affiliation{Department of Chemistry, Princeton University, Princeton, NJ 08544, USA}
\author{Johannes Flick \orcidlink{0000-0003-0273-7797}}
\affiliation{Center for Computational Quantum Physics, Flatiron Institute, 162 5th Avenue, New York, New York 10010,
	USA}
\affiliation{Department of Physics, The City College of New York, 160 Convent Ave, New York, NY 10031}
\affiliation{Department of Physics, The Graduate Center, City University of New York, New York, NY 10016, USA}
\date{\today}
\begin{abstract}
  We present an {\it ab initio} method for computing vibro-polariton and phonon-polariton spectra of molecules and solids coupled to the photon modes of optical cavities.
  We demonstrate that if
	interactions of cavity photon modes with both nuclear and electronic
	degrees of freedom are treated on the level of the cavity Born-Oppenheimer approximation (CBOA),
	spectra can be expressed in terms of the matter response to electric fields and nuclear displacements which are readily available in standard density functional perturbation theory
	(DFPT) implementations.
	In this framework, results over a range of cavity parameters
	can be obtained without the need for additional electronic structure
	calculations, enabling efficient calculations on a wide range of
	parameters. Furthermore, this approach enables results to be more readily
	interpreted in terms of the more familiar cavity-independent molecular electric field response
	properties, such as polarizability and Born effective charges which enter into the vibro-polariton calculation.
	Using corresponding electric field response properties of bulk insulating systems,
	we are also able to obtain $\Gamma$ point phonon-polariton spectra of two dimensional (2D) insulators.
	Results for a selection of cavity-coupled molecular and 2D crystal systems are presented to demonstrate the method. %

\end{abstract}
\maketitle

\section{Introduction}

The coupling of electromagnetic field excitations of optical cavities with
matter has gained recent interest as a means of manipulating material and
molecular properties and processes~\cite{ebbesen2016hybrid,Garcia2021}.
The modification of these
excitations via optical cavities has been reported experimentally for various
characteristics, such as chemical reactivity~\cite{thomas2019tilting}, optical
spectra~\cite{Xiang2020science, PhysRevLett2016Jino}, relaxation dynamics and
ultrafast thermal modification~\cite{APL2021Liu}, intermolecular vibrational
energy redistribution~\cite{science2020Xiang}, enhancement of
ferromagnetism\cite{thomas2021} and thermal control of metal-to-insulator
transition~\cite{Jarc_2023}, among others.

At the same time, these experimental developments have sparked theoretical developments of first principle methods~\cite{FlickRiveraNarang2018,ruggenthaler2023,foley2023,mandal2023}. In the vibrational strong coupling (VSC) regime, where a photonic (cavity) mode becomes resonant with a vibrational/phonon mode, creating vibro-polaritonic states, the so-called cavity Born-Oppenheimer approximation (CBOA)~\cite{flick2017,flick17_cavit_born_oppen_approx_correl} has been applied successfully. %
The CBOA approach treats low-frequency cavity modes, such infrared modes, and nuclei as relatively slow-moving components, in comparison to the fast-moving electrons~\cite{flick17_cavit_born_oppen_approx_correl}. Thus, CBOA treats photons and nuclei equally, allowing for the calculation of adiabatic potential energy surfaces (PES). For the vibrational strong coupling regime, the CBOA has been applied to describe chemical reactivity ~\cite{galego2019,campos2020,Li_2021}, and spectra of vibro-polariton states~\cite{bonini2022} among others~\cite{fiechter2024understanding,Fischer2023,Schnappinger_2023,Schnappinger2023_2,sidler2023unraveling}.

In this article, we present a methodology for obtaining the modification of vibrational modes in
molecular systems and insulating 2D solids by coupling matter to low-frequency photon modes in optical cavities. Results of this method are shown to be equivalent to the linear response framework based on the CBOA~\cite{bonini2022} but use only the first-order electronic response to
electric field and nuclei displacement perturbations as input. Using a mapping of the
CBOA energy functional to a finite field enthalpy, we can also utilize existing
standard {\it ab initio} finite electric field response methods based on the modern
theory of polarization~\cite{King1993, Souza2002} to calculate the cavity
modified phonon spectra in solids.
We show that this
formalism can be generalized to multiple photon modes. Therefore, this mapping
allows us to have clear understanding of the relation between PES inside
and outside the cavity. Additionally, this method yields an improvement in computational efficiency of the techniques previously
developed ~\cite{bonini2022} for the linear response theory of vibro-polaritons in
insulating materials. We present several examples for modulation of
phonons in hexagonal boron nitride (h-BN), $\text{HfS}_{2}$ as a transition metal
dichalcogenide (TMD), and CO$_{2}$ and Fe(CO)$_5$ as two molecular examples to
demonstrate the method.

\begin{figure}
  \includegraphics[width=0.45\linewidth]{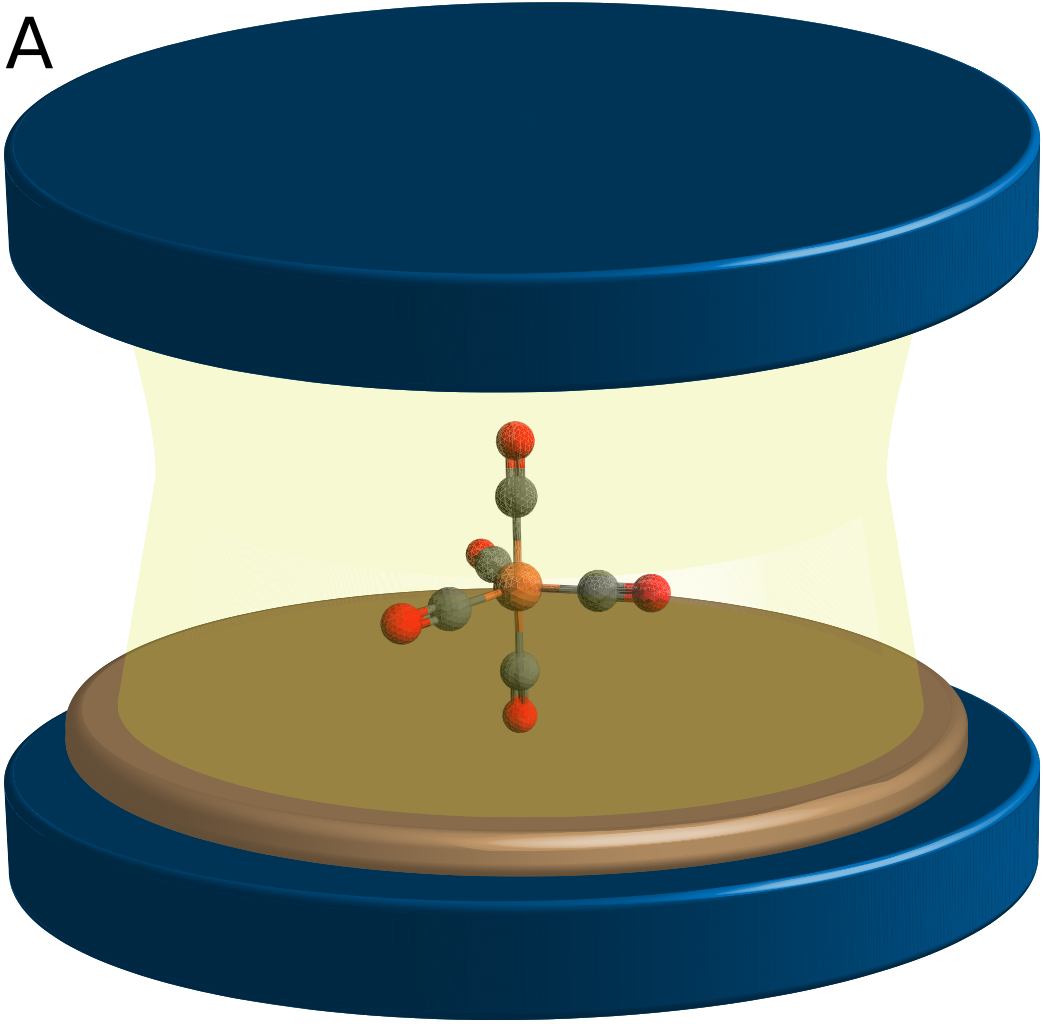}
  \includegraphics[width=0.45\linewidth]{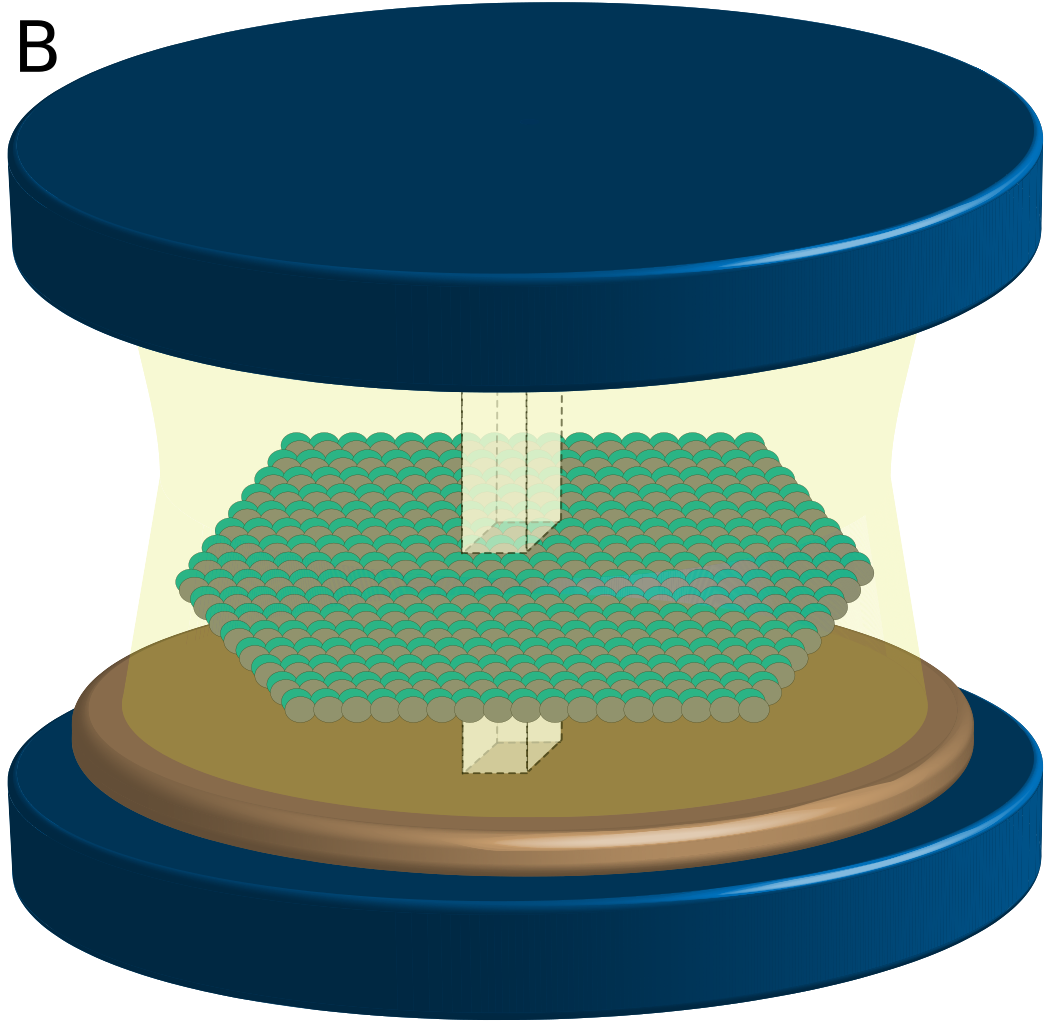}
  \caption{ Schematic illustrations of model matter + optical cavity systems.
	Fig.~A depicts a single molecule coupled to cavity photon modes. Fig.~B
	depicts a mono-layer of a 2D material arranged so that the cavity mirrors
	and the 2D plane of the system are parallel. The parallelepiped passing from
	the upper mirror through the material to the lower mirror represents the
	unit cell of the model used to perform calculations with periodic boundary
	conditions.}
  \label{fig:schematic}
\end{figure}

\section{Formalism \label{sec:formalism}}

\subsection{Mean field CBOA and effective electric fields}

Within the dipole approximation light is assumed to couple to matter 
through the dipole moment $\mu$ of the matter system, which has both nuclear and
electronic contributions, i.e. $\mu = \mu_{e} + \sum_{I}Z_{I}R_{I}$. This is a reasonable assumption for charge-neutral, nonmagnetic systems coupled to photon
fields that are approximately spatially uniform across the matter charge distribution. The minimal coupling Hamiltonian within the length gauge and
dipole approximation is given by
$\hat{H} = \hat{T}_e + \hat{T}_\mathrm{nuc} + \hat{T}_\mathrm{pht} + \hat{V}_\mathrm{Coloumb} + \hat{V}_\mathrm{dipole}$
with~\cite{flick2017,bonini2022}
\begin{equation}
	\label{eq:pt-mu interaction}
	\hat{V}_{\mathrm{dipole}} = \frac{1}{2}\sum_{\alpha}(\omega_{\alpha}\hat{q}_\alpha - \vec{\lambda}_\alpha\cdot\dipoleop)^2 \\.
\end{equation}
where $\hat{q}_{\alpha}$ corresponds to local displacement field of photon mode $\alpha$
with parameters $\omega_{\alpha}$ and $\lambda_{\alpha}$ characterizing the photon mode of the empty
cavity (which is in general dependent on the mirrors and geometry) with
$\omega_{\alpha}$ representing the photon mode frequency and $\lambda_{\alpha}$ being a vector
with direction indicting the mode photon mode polarization, and magnitude
indicating the amplitude of photon field oscillations at the matter
position\cite{selfpolterm}.

In the CBOA~\cite{flick2017,flick17_cavit_born_oppen_approx_correl,bonini2022}, $q_\alpha$ are treated as
``slow variables'' along with nuclei coordinates $R_{I}$. An effective potential
for these slow variables is then constructed from the ground state energies of
the electronic Hamiltonian
$\hat{H}_{e}(R,q) = \hat{T}_e + \hat{V}_\mathrm{Coloumb}(R) + \hat{V}_\mathrm{dipole}(R, q)$ solved at
static coordinates $(R,q)$ yielding the system of equations:
\begin{equation}
	\label{eq:cboa_elec}
	\hat{H}_{e}(R,q) \ket{\elecG(R, q)} = \Fcboaelec(R, q)\ket{\elecG(R, q)}
\end{equation}
\begin{equation}
	( \hat{T}_{\mathrm{nuc}} + \hat{T}_{\mathrm{pht}} + \Fcboa(R, q)) \ket{\Phi^{\mathrm{nuc}, \mathrm{pht}}} = E \ket{\Phi^{\mathrm{nuc}, \mathrm{pht}}}
\end{equation}
where $\Fcboa(R, q)$ corresponds to the solution of Eq.~\ref{eq:cboa_elec} with the
lowest energy $\Fcboaelec(R,q)$ occurring at an electronic ground state $\elecQ(R,q)$.
\footnote{In general the photon and nuclei kinetic
	energy operators can act on the electronic wave function through the
	parametric dependence leading to an effective vector potential on the slow
	variables as well as additional terms in the scalar potential. Since we will
	be investigating dynamics near ground states of non-magnetic systems with an
	electronic gap ($E_{\mathrm{gap}}$), a gauge can be chosen to make the vector
	potential term vanish. Contributions to the scalar potential are on the order
	of $(\frac{\omega \lambda}{E_{\mathrm{gap}}})^2$ which we neglect as we will be
	considering vibrational coupling systems with electronic gaps larger than the
	photon mode frequencies.}

In what follows, we will treat the electronic dipole squared term (dipole-self energy (DSE)) in $V_{\mathrm{dipole}}$ (Eq.~\ref{eq:pt-mu interaction}) with a mean-field approximation, i.e. making the substitution
\begin{equation}
	(\lambda_\alpha \cdot \dipoleop)^2 \stackrel{\text{MF}}{\approx}  2 (\lambda_\alpha \cdot \braket{\dipoleop}) (\lambda_\alpha \cdot \dipoleop) - (\lambda_\alpha \cdot \braket{\dipoleop})^2 \\.
\end{equation}
Where $\braket{\ldots}$ is the expectation value evaluated for the electronic state
at fixed $(R,q)$. To make the function dependence more explicit a dipole
expectation value $\braket{\dipoleop}$ will be written as a function in terms of
the nuclei coordinate and electronic state as $\dipoleelec(\elecG, R)$. Then
$\Fcboa(R,q)$
is found via a self consistent solution to
Eq.~\ref{eq:cboa_elec} where $H_{e}$ is replaced with
\begin{equation}
	\label{eq:mean-field}
	\begin{split}
		&\hat{H}_{e}^{\mathrm{MF}}(\elecG;R, q)  = \hat{T}_e + \hat{V}_\mathrm{Coulomb}(R)                                                                                                                            \\
		                                       & + \sum_{\alpha} \frac{1}{2} (\omega_{\alpha}q_\alpha)^{2} -  (\omega_{\alpha}q_{\alpha}-\lambda_\alpha \cdot \dipoleelec(\elecG,R)) (\lambda_\alpha \cdot \dipoleop) \\
		                                       & - \frac{1}{2}(\lambda_\alpha \cdot \dipoleelec(\elecG,R))^2 .
	\end{split}
\end{equation}
This equation must be solved self-consistently so that $\elecG$ used to
construct the Hamiltonian is equal to the resulting ground state
$\elecQ$.
Self-consistent
electronic solutions can be found by extending existing Kohn-Sham density
functional theory (DFT) to incorporate these terms arising from
$\hat{V}_{\mathrm{dipole}}$ with existing
functionals~\cite{flick17_cavit_born_oppen_approx_correl,flick18_cavit_correl_elect_nuclear_dynam,bonini2022}.

In this work, we develop an alternative, but equivalent procedure where cavity
properties can be obtained using DFT calculations with only finite electric
field terms. We can consider a standard Kohn-Sham DFT functional $\Fclmbelec$
which takes some electronic variables $\psi$ \footnote{$\psi$ may represent the charge
  density or a Kohn-Sham wavefunction, for our purposes these are just the
  electronic variables particular functionals are minimized with respect to and
  from which we can evaluate dipole moments.} along with nuclei coordinates $R$
and yields some energy which is taken as an approximation of the expectation
value of the first two terms in Eq.~\ref{eq:mean-field}
\begin{equation}
	\Fclmbelec(\elecG;R) \approx \braket{\hat{T}_e + \hat{V}_\mathrm{Coulomb}(R)}_{\psi}
\end{equation}
which is minimized for fixed $R$ at $\elecC$ so that one can consider a ground state energy as a function of $R$
\begin{equation}
	\Fclmb(R) = \Fclmbelec(\elecC;R) = \min_{\elecG}\Fclmbelec(\elecG;R).
\end{equation}
Then the inclusion of $\hat{V}_{\mathrm{dipole}}$ terms at a mean-field level corresponds to working with a new functional
\begin{equation}
  \label{eq:CBOA}
	\Fcboaelec(\elecG;R,q) = \Fclmbelec(\elecG;R) + \frac{1}{2}\sum_{\alpha}(\omega_{\alpha}q_{\alpha} - \lambda_{\alpha}\cdot \dipoleelec(\elecG,R))^{2}
\end{equation}
which is minimized for fixed $R,q$ at $\elecQ$ so that one can consider a ground state energy as a function of $R,q$
\begin{equation}
  \label{eq:CBOA_gs}
	\Fcboa(R,q) = \Fcboaelec(\elecQ;R,q) = \min_{\elecG}\Fcboaelec(\elecG;R,q)\\.
\end{equation}

For a given $\elecG$, Eq.~\ref{eq:mean-field} has only one electronic operator
from $V_{\mathrm{dipole}}$ and it is linear in $\dipoleop$. This is much like a
system in a static and uniform electric field. At a self-consistent solution
$\elecQ$ it is as if there is an effective electric field coupling linearly with
the dipole operator given by $\mathbb{E} = \sum_{\alpha}\lambda_{\alpha}\mathcal{E}_{\alpha}$ where
\begin{equation}
	\label{eq:constit_impl}
	\mathcal{E}_\alpha = \omega_\alpha q_\alpha - \lambda_\alpha \cdot \dipoleQ(R, q)\\
\end{equation}
and $\dipoleQ(R, q) = \dipoleelec(\elecQ; R,q)$. Since $\elecQ$ is a fixed point
in a procedure for finding self-consistent solutions to Eq.~\ref{eq:mean-field}
the terms corresponding to expectation values act as constants.
So long as there is not a lower
energy state for the same effective electric field (e.g. as can occur in ferroelectric
materials in regions where
$\partial^{2}\Fcboa/\partial q^{2} < 0$)
then
the resulting electronic ground state $\elecQ$
will also be a fixed point of a procedure
self consistently minimizing the enthalpy functional
\begin{equation}
	\label{eq:enthalpy}
	\Fefieldelec(\elecG;R,\mathcal{E}) = \Fclmbelec(\elecG;R) - \mathbb{E}(\mathcal{E})\cdot \dipoleelec(\elecG, R_{I})\\.
\end{equation}
This functional is minimized for fixed $R,\mathcal{E}$ at $\elecE$ so that one can consider a ground state enthalpy as a function of $R,\mathcal{E}$
\begin{equation}
	\label{eq:enthalpy_gs}
	\Fefield(R,\mathcal{E}) = \Fefieldelec(\elecE;R,q) = \min_{\elecG}\Fefieldelec(\elecG;R,\mathcal{E})\\.
\end{equation}
Self-consistent minimization of this functional is already implemented in many
DFT codes either as a linear external potential or via a Berry phase formalism
to treat insulating systems with periodic boundary conditions\cite{Souza2002}.

  The relation between Eq.~\ref{eq:CBOA} and Eq.~\ref{eq:enthalpy_gs} is
quite similar to the relation between static displacement field and static
electric field functionals identified in Ref.~\cite{Stengel_2009}, which was
presented in the context of bulk dielectrics. Analogously to that work we can
derive the relation between Eq.~\ref{eq:CBOA} and Eq.~\ref{eq:enthalpy} in terms of the
Legendre transform (LT) of Eq.~\ref{eq:CBOA}. We start by defining $\mathcal{E}_{\alpha}$ to be the
conjugate variable to the quantity $\mathcal{D}_{\alpha} := \omega_{\alpha}q_{\alpha}$. Then so long as
Eq.~\ref{eq:CBOA} is convex with respect to $q_{\alpha}$ the LT of
$\Fcboaelec(\elecG, R, d)$, which we label as $\mathfrak{F}(\elecG, R, \mathcal{E})$ is then given by
\begin{equation}
  \mathfrak{F} := \Fcboaelec - \sum_{\alpha}\mathcal{D}_{\alpha}\mathcal{E}_{\alpha}
\end{equation}
where

\begin{minipage}{0.229\textwidth}
\begin{equation}
  \label{eq:edual}
  \mathcal{E}_{\alpha} := \frac{\partial \Fcboaelec}{\partial \mathcal{D}_{\alpha}}
\end{equation}
\end{minipage}
\begin{minipage}{0.229\textwidth}
\begin{equation}
  \label{eq:ddual}
  \mathcal{D}_{\alpha} := -\frac{\partial \mathfrak{F}}{\partial \mathcal{E}_{\alpha}}
\end{equation}
\end{minipage}
where Eq.~\ref{eq:edual} provides a systematic means of arriving at
Eq.\ref{eq:constit_impl}.
The LT of $\Fcboaelec$ written in terms of $\mathcal{E}_{\alpha}$ is then
\begin{equation}
  \label{eq:directLT}
  \mathfrak{F}(\elecG, R, \mathcal{E}) = \Fclmb(\elecG, R) - \sum_{\alpha} \mathcal{E}_{\alpha} \lambda_{\alpha} \cdot \dipoleE(\elecG, R) - \frac{1}{2}\sum_{\alpha}\mathcal{E}_{\alpha}^{2} \\.
\end{equation}
This function differs from $\Fefieldelec$ only by the term
$- \frac{1}{2}\sum_{\alpha}\mathcal{E}_{\alpha}^{2}$, which is constant for fixed $\mathcal{E}$, and thus
minimizing $F$ at fixed $\mathcal{E}$ with respect to $\elecG$ would yield the same
$\elecE$ as defined in Eq.~\ref{eq:enthalpy_gs}.

In terms of differentials, we have
\begin{equation}
  d\Fcboaelec = \frac{\partial \Fcboaelec}{\partial \elecG}d\elecG+ \frac{\partial \Fcboaelec}{\partial R}dR + \mathcal{E} d\mathcal{D}
\end{equation}
and using Eq.~\ref{eq:ddual}
\begin{equation}
  d\mathfrak{F} = \frac{\partial \mathfrak{F}}{\partial \elecG}d\elecG + \frac{\partial \mathfrak{F}}{\partial R}dR - \mathcal{D} d\mathcal{E}
\end{equation}
Since $\mathfrak{F} := \Fcboaelec - \sum_{\alpha}\mathcal{D}_{\alpha}\mathcal{E}_{\alpha}$ we also have
\begin{equation}
  d\mathfrak{F} = d\Fcboaelec -  \mathcal{E} d\mathcal{D} - \mathcal{D} d\mathcal{E}
\end{equation}
so that
\begin{equation}
  d\mathfrak{F} = \frac{\partial \Fcboaelec}{\partial \elecG}d\elecG+ \frac{\partial \Fcboaelec}{\partial R}dR - \mathcal{D} d\mathcal{E} \\.
\end{equation}
Since $\frac{d \mathfrak{F}}{d \psi}\vert_{R \mathcal{E}} = \frac{d \Fcboaelec}{d \psi}\vert_{R \mathcal{D}}$ the electronic
state $\elecE$ which minimizes $\mathfrak{F}$ (and $\Fefieldelec$) at fixed $R, \mathcal{E}$ will
also minimize $\Fcboaelec$ at fixed $R, \mathcal{D}$.
At the electronic ground state $\frac{\partial U}{\partial \psi}=0$ so the force $F_I$ on the $I$-th nuclei is given by
\begin{equation}
  F_I = -\frac{d \Fcboa}{d R_{I}}|_{\mathcal{D}} = -\frac{d \Fefield}{d R_{I}}|_{\mathcal{E}}
\end{equation}
thus forces on nuclei computed as derivatives of $\Fefield$ with respect to
$R_{I}$ are equal to those obtained from the corresponding derivatives of $\Fcboa$.
The effective force $F_\alpha$ on photon mode variable $q_{\alpha}$ is
\begin{equation}
  F_\alpha = -\frac{d \Fcboa}{d q_{\alpha}}\vert_{R} = -\omega_{\alpha}\frac{d \Fcboa}{d \mathcal{D}_{\alpha}}\vert_{R} = -\omega_{\alpha} \mathcal{E}_{\alpha}
\end{equation}
where Eq.~\ref{eq:edual} was used to arrive at the last equality.

 It can be advantageous to compute properties of light-matter
coupled systems by transforming results obtained via Eq.~\ref{eq:enthalpy}
rather than using a direct implementation of Eq.~\ref{eq:mean-field}. Not only
is Eq.~\ref{eq:enthalpy} already integrated into many first principles codes,
but also cavity parameters such as $\lambda_{\alpha}, \omega_{\alpha}$ enter only in the transformation, implemented as a computationally cheap post-processing step,
	so that a range of such parameters can be obtained from an initial set of
	first-principles calculations. However, while it is simple to obtain the
electric field parameters $\mathcal{E}_{\alpha}(R, q)$ from a known $\dipoleQ(R, q)$, this
function can not be obtained directly from electric field calculations where one
instead can only access $\dipoleE(R, \mathcal{E}) = \dipoleelec(\elecE, R, \mathcal{E})$. It is only
when $R,q,\mathcal{E}$ are such that Eq.~\ref{eq:constit_impl} is satisfied that
$\dipoleE(R, \mathcal{E}) = \dipoleQ(R,q)$.

In general finding the electric
field parameters yielding the same state as $(R, q)$ requires solving the system
of equations
\begin{equation}
	\label{eq:recursive}
	\mathcal{E}_{\alpha}(R, q) = \omega_{\alpha} q_{\alpha} - \lambda_{\alpha} \cdot \dipoleE(R, \sum_{\beta}\lambda_{\beta}\mathcal{E}_{\beta}(R, q))
\end{equation}
for each function $\mathcal{E}_{\alpha}(R, q)$. If such a relation is solved it
can then be used to obtain $\dipoleQ(R,q)$, $\Fcboa(R, q)$ and
other properties of the light-matter coupled system by post-processing results
computed for the matter at static electric fields.

\subsection{Linear response framework}
While in general solving Eq.~\ref{eq:recursive} is non-trivial, properties such
as the vibro-polariton spectra can be obtained using only linear terms in
$\dipoleE(R, \mathcal{E})$ so that Eq.~\ref{eq:recursive} becomes a linear problem. In
this section we show how vibro-polariton modes and IR spectra can be obtained
from standard adiabatic response terms, namely the force constant matrix, Born
effective charges, and clamped ion polarizability. Near the ground state values
of $R$ and $\mathcal{E}$ (and thus near $R$ and $q$) the dipole expectation value can be
expanded to linear order
\begin{equation}
	\label{eq:lin-mu}
	\dipoleM = \dipoleM_{0} + \sum_{I}Z^{*}_{I}\Delta R_{I} + \chi \mathbb{E}
\end{equation}
where $Z_{I}^{*}$ are Born effective charges
($Z_{Ii} = \frac{d\dipoleM_{i}}{d R_{I}}$) and $\chi$ is the
polarizability tensor ($\chi_{ij} = \frac{d \dipoleM_{i}}{d\mathbb{E}_{j}}$).

Then Eq.~\ref{eq:recursive} becomes
\begin{equation}
	\mathcal{E}_{\alpha} = \omega_{\alpha} \Delta q_{\alpha} - \lambda_{\alpha}\cdot \sum_{I}Z^{*}_{I}\Delta R_{I} - \lambda_{\alpha}^{T}  \chi \sum_{\beta} \lambda_{\beta} \mathcal{E}_{\beta}
\end{equation}
where $\Delta q_{\alpha} = q_{\alpha} - \frac{\lambda_{\alpha}\cdot \dipoleM_{0}}{\omega_{\alpha}}$.
This linear system can then be solved for $\mathcal{E}_\alpha$
\begin{equation}
	\label{eq:efield_q_reln}
	\mathcal{E}_{\alpha} = \sum_{\beta}[(\mathbb{I} + \mathbb{X})^{-1}]_{\alpha \beta} (\omega_{\beta} \Delta q_{\beta} - \lambda_{\beta}\cdot \sum_{I}Z^{*}_{I}\Delta R_{I})
\end{equation}
where $\mathbb{I}_{\alpha \beta} = \delta_{\alpha \beta}$ is an identity matrix and $\mathbb{X}_{\alpha \beta} = \lambda_{\alpha}^{T}\chi\lambda_{\beta}$.
Electronic ground states found by minimizing energy via Eq.~\ref{eq:mean-field} at small, fixed,
changes in $q$ from the $\lambda \rightarrow 0$ ground state value are then identical to states found via fixed electric field calculations with $\mathbb{E} = \sum_{\alpha}\lambda_{\alpha}\mathcal{E}_{\alpha}$ using Eq.~\ref{eq:enthalpy} with $\mathcal{E}_{\alpha}$ given by Eq.~\ref{eq:efield_q_reln}.

We next construct second order expansion of the energy $\Fcboa$ in
both nuclei coordinates and electric field terms, then use this linear
relation (Eq.~\ref{eq:efield_q_reln}) to perform a change of coordinates from
the $(\mathcal{E}_{\alpha}, R_{I})$ basis to the desired $(q_{\alpha}, R_{I})$ basis. Note that
while we have defined the adiabatic combined light-matter energy $\Fcboa$ as
one computed self consistently at fixed $(R_{I},q_{\alpha})$, we can nevertheless use
Eq.~\ref{eq:constit_impl} to express this energy as
\begin{equation}
	\Fcboa(R,\mathcal{E}) = \Fclmbelec(\elecE;R) + \frac{1}{2}\sum_{\alpha}\mathcal{E}_{\alpha}^{2}\\.
\end{equation}
We can obtain the (total) derivative of $\Fclmbelec$ with respect to $\mathcal{E}_{\alpha}$ by
applying the variation Hellman-Feynman theorem to the $\Fefield$ functional, which
states that
$d \Fefield/d\mathcal{E_{\alpha}} = \partial \Fefield/\partial \mathcal{E_{\alpha}}$, thus
\begin{equation}
	\frac{d \Fclmbelec(\elecE;R)}{d \mathcal{E}_{\alpha}} = \sum_{\beta}\mathcal{E}_{\beta} \lambda_{\beta} (\chi \lambda_{\alpha} + \frac{d \chi}{d \mathcal{E}_{\alpha}}\sum_{\gamma}\mathcal{E}_{\gamma}\lambda_{\gamma}) \\. \end{equation}
We focus on regimes near zero electric field and only expand energy to second
order so the second term may be dropped.
At zero electric field derivatives with nuclei are unchanged from the
usual force constant matrix given by
$C_{IJ} = \frac{\partial^{2}\Fcboa}{\partial R_{I} \partial R_{J}}$.
Then the energy $\Fcboa$ to second order in $R,\mathcal{E}$ variables is given by
\begin{equation}
	\label{eq:model-efilds}
	\begin{split}
		\Fcboa & =  \Fcboa(R_{0}, q_{0}) + \frac{1}{2}C_{IJ}\Delta R_I\Delta R_J                                                    \\
		       & + \frac{1}{2}(\delta_{\alpha\beta} + \lambda_{\beta} \chi \lambda_{\alpha})\mathcal{E}_{\alpha}\mathcal{E}_{\beta}
	\end{split}
\end{equation}

Using Eq.~\ref{eq:efield_q_reln} this can be transformed to finally obtain the second order energy in the $R_{I},q_{\alpha}$ coordinates.
\begin{widetext}
\begin{equation}
	\label{eq:model}
		\Fcboa  =  \Fcboa(R_{0}, q_{0}) + \frac{1}{2}\Delta R_IC_{IJ}\Delta R_J
		        + \frac{1}{2}
		[\omega_\alpha \Delta q_\alpha - \lambda_\alpha Z_I \Delta R_I]
		\left[(
			\mathbb{I} + \mathbb{X}
			\right)^{-1}]_{\alpha \gamma}[\omega_\gamma \Delta q_\gamma - \lambda_\gamma Z_J \Delta R_J]
\end{equation}
\end{widetext}
Eq.~\ref{eq:model} represents a key
result of this work, with the CBOA energy expressed to second order in nuclei
displacements $\Delta R_{I}$ and cavity photon displacements $q_{\alpha}$ using quantities
involving only the systems response to nuclei displacements and static electric
fields.

As in Ref.~\cite{bonini2022} we can obtain IR
absorption amplitudes from the normal mode eigenvectors and the derivative of
the polarization with respect to each degree of freedom. Note that these
quantities represent photon absorption into matter degrees of freedom through
changes in the dipole moment both via displacements of nuclei as well as purely
electronic changes obtained using the adiabatic electronic polarizability
($\chi$). This is not to be confused with the macroscopic absorption by the
entire cavity-matter system which includes the absorption of incident photons into
confined cavity modes.

To compute the IR response amplitude we require an expression for
$\dipoleM(\{q\},\{R\})$ to linear order in all $q_{\alpha}$ and $R_{I}$. Combining
Eq.~\ref{eq:constit_impl} and Eq.~\ref{eq:lin-mu}
\begin{equation}
	\dipoleM = \dipoleM_0 + \chi [\sum_\alpha \vec{\lambda}_\alpha(\omega_\alpha q_\alpha - \vec{\lambda}_\alpha \cdot \mu)] + \sum_i \vec{Z}_i^* \Delta R_i
\end{equation}

with $\dipoleM = \dipoleM_{0} + \Delta \dipoleM$
\begin{equation}
	\Delta \dipoleM + \sum_{\alpha}(\chi \lambda_{\alpha}) \vec{\lambda}_{\alpha}\cdot \Delta \dipoleM = \chi \sum_\alpha \vec{\lambda}_\alpha\omega_\alpha \Delta q_\alpha + \sum_i \vec{Z}_i^* \Delta R_i
\end{equation}
One can compute the $3\times3$ matrix $\Lambda = 1 + \sum_{\alpha} \chi \lambda_{\alpha}\lambda_{\alpha}^{T}$ (where the transpose indicates this is an outer product) then
\begin{equation}
	\label{eq:ir}
	\Delta \dipoleM = \Lambda^{-1} (\chi \sum_\alpha \vec{\lambda}_\alpha\omega_\alpha \Delta q_\alpha + \sum_i \vec{Z}_i^* \Delta R_i) \\.
\end{equation}
The IR amplitude for a given polariton mode is then proportional to the
magnitude of $\Delta \dipoleM$ given by \ref{eq:ir} using the $\Delta q_{\alpha}$ and
$\Delta R_{i}$ eigen-displacements corresponding to that mode.

\subsection{Extension to 2D periodic systems}

To date, there have been only a handful of first principles based investigations
of crystal systems coupled to cavities
\cite{latini2021ferroelectric,latini2019cavity,lu2024cavityengineered} in part
due to several challenges not faced in molecules.
One challenge is that periodic
systems coupled to arbitrary cavity photon modes necessitate a beyond dipole
approximation treatment. Another is that working with an electronic position
operator (and consequently dipole operator) requires some care in the context of
periodic boundary conditions. While we do not address the former issue beyond
only choosing to couple to modes that have a uniform field across the entire
crystal, the formulation of the problem in terms of electronic responses to
electric fields is useful in addressing the latter. Eq.~\ref{eq:model} relies
only on the systems dielectric susceptibility, Born effective charges, and force
constant matrix; all of these properties are routinely computed for periodic
solids. However, some care must be taken in adapting Eq.~\ref{eq:model} to
solids, as we are limited to cavity-matter geometries compatible with the
underlying dipole approximation and must work with intrinsic properties of the
light-matter system while Eq.~\ref{eq:model} was arrived at in the previous
sections while working with the extrinsic properties of dipole moment implicitly
cavity mode ``coupling strength''.

For our treatment of extended solids to be valid in the context of the dipole
approximation we require a cavity-matter geometry where the cavity photon mode
profile is approximately constant over the matter (leading to a spatially
uniform coupling strength). We choose a geometry of a 2D system aligned parallel
and placed between two planar mirrors. Then the confined zero momentum photon
modes couple uniformly to in-plane matter dipole fluctuations. To handle
periodic boundary conditions we reformulate the problem to one in which we deal
with an energy per unit cell given in terms of intrinsic properties of the
system. Thus we consider the case where the entire cavity-matter system is
periodic in two dimensions with lattice vectors equal to those of the 2D unit
cell of the crystal. This sort of 2D periodic cavity+matter unit cell is
illustrated in Fig.~\ref{fig:schematic}B. This geometry allows for a recovery of
Bloch's theorem which can be applied to both the matter as well as the cavity
modes. Given that we are still specialized to the dipole approximation we then
treat only zero momentum ($\Gamma$ point) excitations as beyond this point the cavity
portion modes have spatially modulated coupling strength over the crystal.

Then the energy $\Fcboaelec$ can be considered an energy per unit cell if the
dipole $\dipole$ is taken to be a dipole per unit cell and the definition of
photon displacement $q_{\alpha}$ and coupling strength $\lambda_{\alpha}$ are also redefined for
this periodic context. This can be achieved with the following substitutions for
number of unit cells $N$: $\mu \rightarrow \mu/N$, $q_{\alpha} \rightarrow q_{\alpha} / \sqrt{N}$, and
$\lambda_{\alpha} \rightarrow \lambda_{\alpha}\sqrt{N}$. One interpretation of this formulation is that
the zero momentum cavity eigenstates are normalized over the periodic
cavity-matter cell rather than over all space. Once these substitutions are made
and the parameters of electronic susceptibility tensor $\chi$ and Born effective
charge $Z^{*}$ are given as derivatives of the matter dipole per unit cell
rather than derivatives of the total dipole the application of
Eq.~\ref{eq:model} is essentially unchanged; the energy is now interpreted as an
energy per unit cell.

\section{Methods}

We compute the vibro-polariton spectra for two molecules (CO$_{2}$ and
Fe(CO)$_{5}$) and the $\Gamma$-point phonon-polariton spectra for two 2D solids
(BN and HfS$_{2}$) each with various cavity parameters ($\omega_{\alpha}$ and
$\lambda_{\alpha}$) using the formalism developed in Sec.~\ref{sec:formalism}. For
each system we focus on cavity parameters with a cavity photon mode in resonance
with the vibration mode with the largest magnitude mode effective charge
($Z^{*}_{\mathrm{res}}$, the change in dipole with respect to vibration mode
amplitude outside the cavity) where the photon mode polarization
($\lambda_{\mathrm{res}}$) is aligned with $Z^{*}_{\mathrm{res}}$. For each
system we compare the $|\lambda_{\mathrm{res}}|$ dependence of Rabi splitting and matter IR absorption with and without the inclusion of the
adiabatic electronic polarizability ($\chi$) for both the cases of a single
resonant cavity mode, as well as with the inclusion of additional off-resonant
cavity mode harmonics. In the latter case, we choose the third relevant cavity
harmonic to be in resonance with the matter system to capture the increased
photon density of states at higher frequencies so that
$\lambda_{\alpha} = \frac{\alpha}{3}\lambda_{\mathrm{res}}$ and
$\omega_{\alpha} = \frac{\alpha}{3}\omega_{\mathrm{res}}$. For the two solid
systems each in plane IR active phonon mode.

Matter vibrational modes, polarizabilities, and mode effective charges are
computed via density functional perturbation theory using the VASP density
functional theory code\footnote{Please note commercial software is identified to specify procedures. Such identification does not imply recommendation by National Institute of Standards and Technology (NIST)}\cite{Blochl1994, Kresse1999, Kresse1996,
  Kresse1993} with the PBE
functional\cite{perdew1981} and an energy cutoff of 520 eV. Calculations for
molecular systems were performed as $\Gamma$ point only calculations in a periodic
box with at least 17{\AA}  of vacuum between periodic images along each direction.
Calculations for 2D solid systems were performed on an $8\times8\times1$ Monkhorst-Pack
grid\cite{PhysRevB.13.5188} where the third dimension (orthogonal to the plane
of the material) is treated as periodic with a vacuum of at least 23\AA \,
between periodic images. The most relevant matter properties are tabulated in
Tab.~\ref{tab:matter}. %

\begin{table}
  \caption{\label{tab:matter} Selected matter properties. Units for the
	polarizability element $\chi_{11}$ are given in Bohr radii cubed ($a_{0}^{3}$)
	for molecular systems and $a_{0}^{3}$ per unit cell (u.c.) for 2D crystal
	systems.}
	\begin{tabular}{l|rrcr}
		System     & $\omega_\mathrm{res}$ (meV) & $\chi_{11}$ ($a_{0}^{3}$) & $\vert Z^*_\mathrm{res} \vert$ $(\frac{e}{\sqrt{\mathrm{amu}}})$ & Degenerate \\
		\hline
		CO$_2$     & 312                         & 26                    & 0.75                                                             & False      \\
		Fe(CO)$_5$ & 252                         & 115                   & 0.60                                                             & False      \\
		h-BN       & 165                         & 37  / u.c.                 & 1.10                                                             & True       \\
		HfS$_2$    & 18                          & 316 / u.c.                 & 0.98                                                             & True       \\
	\end{tabular}
\end{table}

\section{Results and Discussion}

We apply the developed methodology to two molecular systems (CO$_{2}$ and
Fe(CO)$_{5}$) and two 2D insulators (BN and HfS$_{2}$).
Vibro-polariton spectra computed by this method are, as expected,
equivalent to those obtained via explicitly incorporating cavity parameters in
the electronic structure calculation as was done in Refs.\cite{bonini2022, flick18_cavit_correl_elect_nuclear_dynam}.
A direct comparison for the case of
a CO$_{2}$ molecule coupled to a cavity mode computed with the Octopus DFT code\cite{octopus2020} is
attached as supplemental material \sm{for a comparison with explicitly including $\hat{V}_\mathrm{dipole}$ in the DFT energy functional.}.
For each system we
present vibro-polariton mode information for a cavity with a photon mode in
resonance with the phonon that has the largest IR absorption amplitude.
Results are presented for calculations where this resonant cavity mode is the only
photon mode included in the calculation (labeled as
$N_{\alpha}=1, \alpha_{\mathrm{res}}=1$) as well as for the case where 7 cavity
harmonics are included with the third harmonic being the resonant mode(labeled
as $N_{\alpha}=7, \alpha_{\mathrm{res}}=3$). Each of these results are presented
with and without the inclusion of the adiabatic electronic interaction; plots
which artificially neglect this interaction are indicated with the label ``$\chi \rightarrow 0$''.

\begin{figure}
	\centering
	\includegraphics[width=\linewidth]{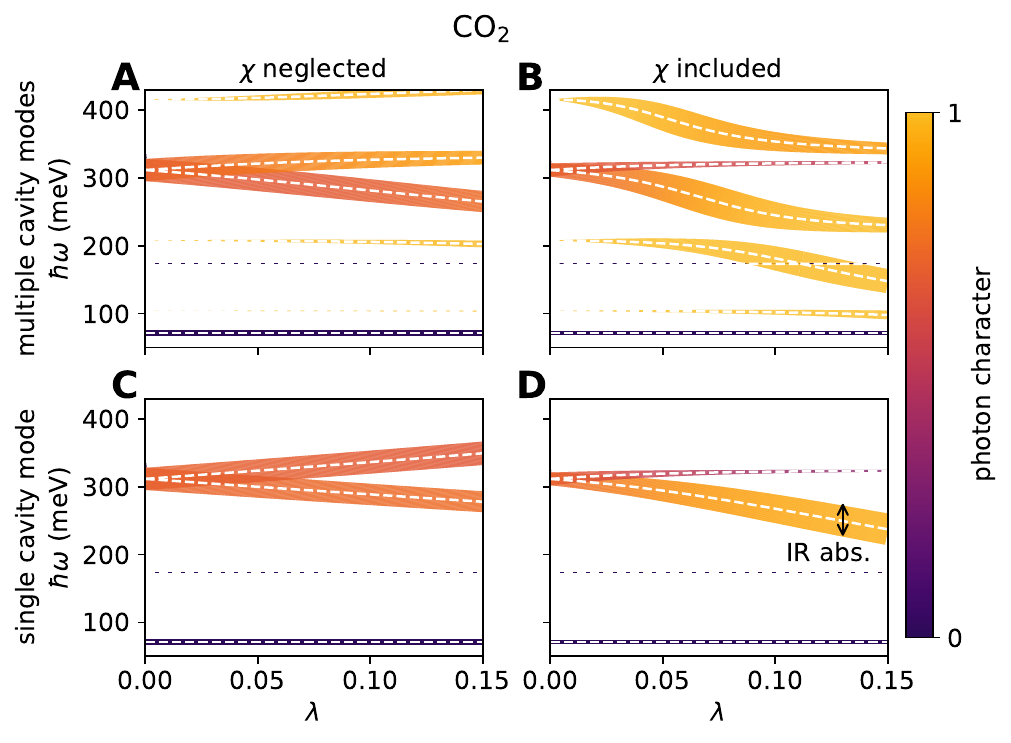}
	\caption{\label{fig:co2} Computed vibro-polariton spectrum vs coupling strength ($\lambda$)
		for CO$_{2}$.
		The thickness of curves is proportional to the IR absorption,
		while the color indicates the sum of the absolute values of photon
		components of the mode eigenvector. The right plots include the electronic
		polarizability term of Eqs.~\ref{eq:model} and \ref{eq:ir}, while on the
		left the value is set to zero. The spectra reveal that electronic polarizability significantly influences the strong regime of light-matter coupling, as depicted. Moreover, electronic polarizability can impact both the IR absorption and the photon components of the mode eigenvector. The lower plots indicate calculations done with a single photon mode, while in the upper two plots, 7 photon mode
		harmonics are included with the third photon mode in resonance with an IR active vibrational mode. In these upper two plots, coupling strength is scaled linearly with mode frequency and the value on the horizontal axis
		indicates the coupling strength of this third resonant mode.}
\end{figure}

Results for CO$_{2}$ with a cavity mode frequency
$\omega_{\mathrm{res}} \approx 300\; \text{meV}/\hbar$ in resonant with the $A_{u}$
vibration mode are shown in Fig.~\ref{fig:co2}. Out of the cases investigated in
this work this system has the closest resemblance to simple models involving a
single matter mode ``emitter'' as the relevant IR active vibration mode is well
separated in energy from other degrees vibration modes. Indeed, Fig.~\ref{fig:co2}C shows results for only a single resonant photon mode with the adiabatic
electronic response ($\chi$) is artificially neglected yielding the standard Rabi
splitting behavior with upper and lower polaritons having similar photon
character, IR response, and frequencies which change essentially linearly with
coupling strength ($\lambda$) with slopes of opposite sign, but equal magnitude.

In contrast when the electronic response is incorporated even the case of a
single photon yields asymmetries between upper and lower polaritons which are seen in
the Fig.~\ref{fig:co2}D. This
asymmetric energy splitting arises largely from the $\lambda$
dependence of the $\mathbb{X}$ term in Eq.~\ref{eq:model}. If we were to
decouple the photon mode from the nuclei vibration but not the electronic
response, the photonic mode frequency would change as
$\omega(\lambda) = \omega(0)/\sqrt{1 + \lambda^{2}\chi}$. Thus as $\lambda$
increases the photon mode is now partially in the ``dielectric medium'' of the
electronic state leading to a change in frequency, effectively detuning the
photon and vibration modes.
This results in the nonlinear asymmetric splitting
where the upper mode polariton only increases slightly in frequency as in the large $\lambda$ regime its
character resembles more of a pure vibration mode. The
lower polariton decreases in frequency at small $\lambda$
similarly to the Rabi splitting case, however at large $\lambda$ the changes are driven
more by the renormalization of the photon frequency by the electronic response.
In addition, the inclusion of the electronic response also leads to an
additional contribution to the IR amplitude; changes in photon $q$ now lead to
changes in matter dipole through the electronic response. The upper polariton
mode involves $q$ and $\Delta R$ which favor oppositely oriented dipole moments
thus the nuclear and electronic changes in dipole moments are in opposite
directions leading to a smaller overall IR amplitude in the upper polariton.
The lower polariton however involves photon and nuclear displacements which
favor constructive changes in dipole and thus has a larger peak. Furthermore,
the electronic change in dipole due to changes in $q$ increases with coupling strength leading to an increase in IR absorption of the lower polariton as $\lambda$ increases.
While the $\Delta R^{2}$ term
can also yield asymmetry this term is included in the $\chi\rightarrow 0$ results
and is seen to have a much smaller effect compared to the electronic response.

Fig.~\ref{fig:co2}A and B show the vibro-polariton spectra when multiple cavity
harmonics are included in the calculation. Fig.~\ref{fig:co2}A shows results 
in the absence of the adiabatic electronic response ($\chi \rightarrow 0$) where there is some
mixing of polariton states with initially off-resonant cavity harmonics only at
very strong coupling strength. Otherwise, this picture is not so different than
the single cavity mode case in the region around the resonant vibration.
However, when the electronic response is included (Fig.~\ref{fig:co2}B) the
spectra is modified in several ways. Since the electronic system responds
(adiabatically) to photon displacements even off-resonant modes acquire an IR
absorption strength and have their frequency renormalized as was the case with a
single photon mode. Furthermore, once the photon modes are dressed by the
electronic response they are no longer non-interacting even in the absence of
interactions with nuclear motion as can be seen in the coupling term between
$\Delta q_{\alpha}$ and $\Delta q_{\gamma}$ in Eq.~\ref{eq:model} which leads to some changes in
frequency even between modes that involve no nuclear motion.

\begin{figure}
	\centering
	\includegraphics[width=\linewidth]{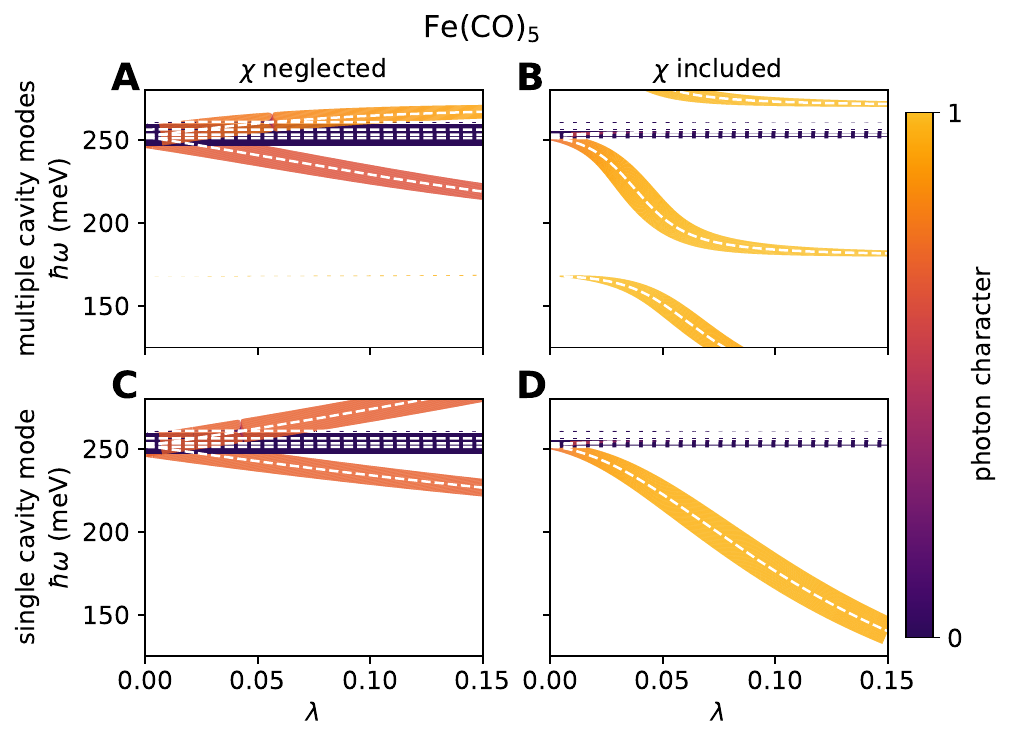}
	\caption{\label{fig:feco5} Computed vibro-polariton spectrum vs coupling strength ($\lambda$)
		for Fe(CO)$_{5}$.
		See the caption for Fig.~\ref{fig:co2} for details on interpreting the plot.
		The upper right
		plot contains an inset with the vertical axis scaled to better see the
		$\lambda$ dependence in the shown region. As observed here, the electronic polarizability can impact the spectra, IR absorption, and the photon components of the mode eigenvector.}
\end{figure}

\begin{figure}
	\centering
	\includegraphics[width=\linewidth]{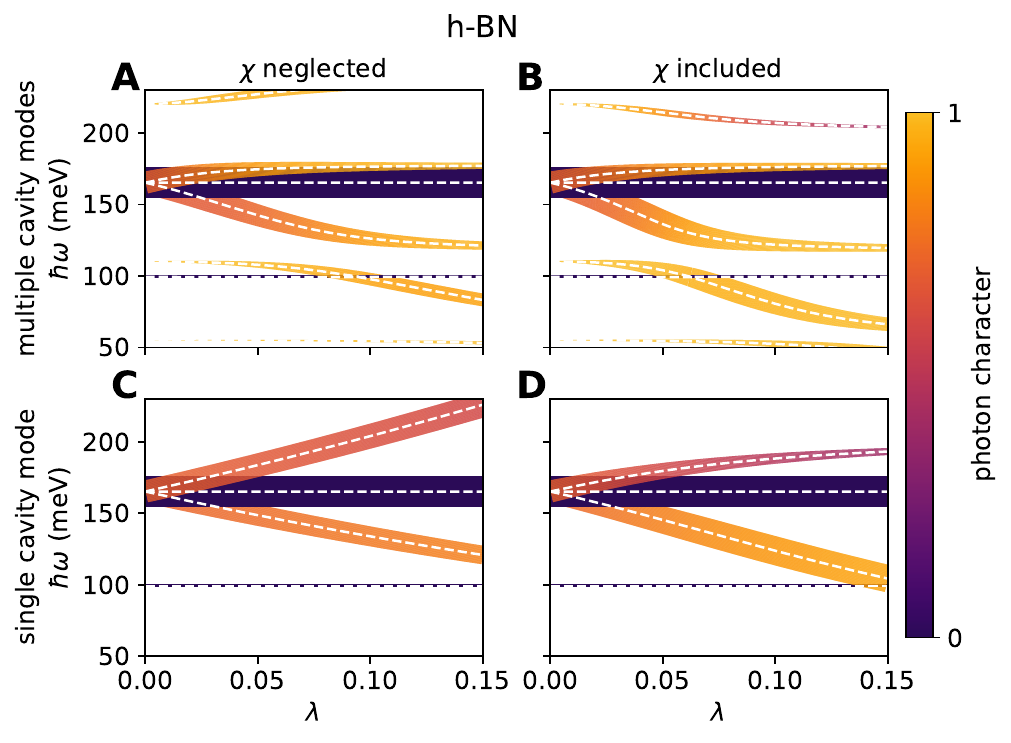}
	\caption{\label{fig:hbn}
		Computed vibro-polariton spectrum vs coupling strength ($\lambda$)
		for h-BN as a crystalline material.
		See the caption for Fig.~\ref{fig:co2} for details on the plot interpretation.}
\end{figure}

\begin{figure}
	\centering
	\includegraphics[width=\linewidth]{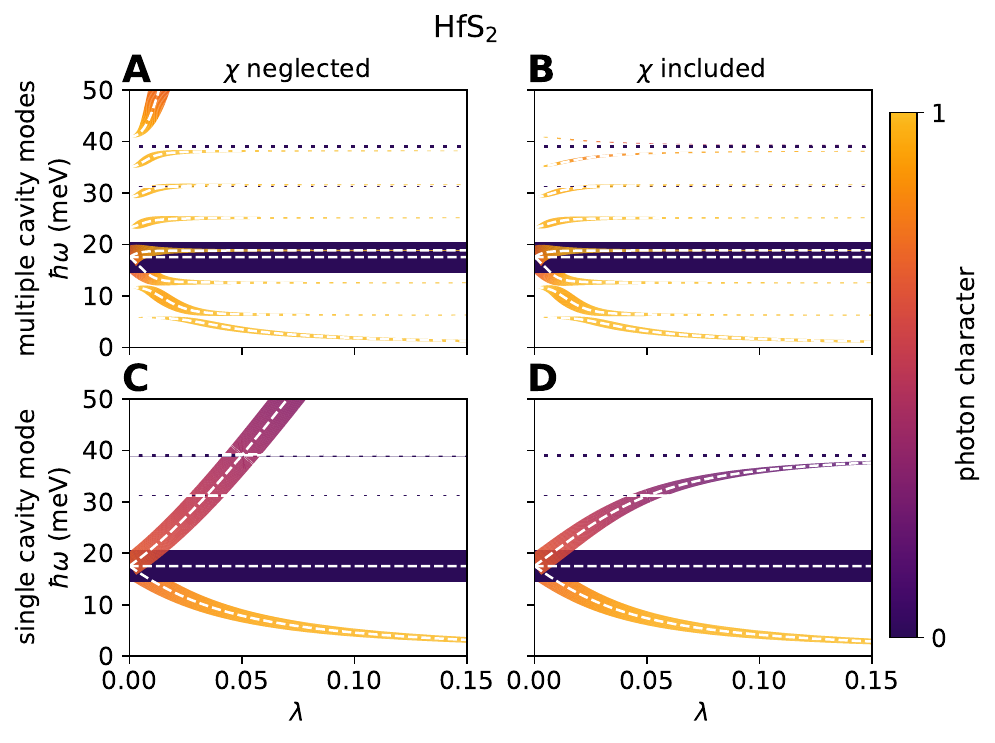}
	\caption{\label{fig:hfs2}Computed vibro-polariton spectrum vs coupling strength ($\lambda$)
		for HfS$_{2}$.
		Refer to the caption for Fig.~\ref{fig:co2} for details regarding the interpretation of the plot. As shown here, the very sharp Rabbi splitting of the lowest vibrational mode show in the plot is moderated in the strong coupling regime by including the electronic polarizability in the calculation of the spectra.}
\end{figure}

These same effects can also be observed observed in the cases of Fe(CO)$_5$ in
Fig.~\ref{fig:feco5}, h-BN in Fig.~\ref{fig:hbn}, and HfS$_2$ in
Fig.~\ref{fig:hfs2} though the spectra of these systems contain a more
complicated set of features than CO$_{2}$. Differences between how these effects
manifest in the vibro-polariton spectra of different systems can largely be
understood in terms of each systems (free-space) vibrational spectra, electric
polarizability ($\chi$), and mode effective charges $Z^{*}$. These quantities, all
of which can be computed with standard (non-QED) density functional perturbation
theory, are tabulated in Table.\ref{tab:matter}.

As can be seen in Fig~\ref{fig:feco5}, Fe(CO)$_5$ exhibits a pronounced shift of
lower polaritons towards lower energies, as depicted in Fig.~\ref{fig:feco5},
when the polarizability is incorporated into the vibro-polariton calculation.
This is largely due to the relevant component of the polarizability $\chi$ being
roughly four times the magnitude as in CO$_{2}$. As in CO$_{2}$, the dipole
moments of the system exhibit constructive changes for $\Delta R$ and $q$ in the
lower polaritons, while in the upper polaritons, these changes are destructive.
Consequently, incorporating the polarizability $\chi$ leads to a reduction in the
IR response in the upper polaritons and an enhancement in the lower polaritons.
Since $\chi$ is larger in Fe(CO)$_5$ this effect sets in more rapidly with respect
to increases in $\lambda$ than in CO$_{2}$. The frequency region near resonance in
Fe(CO)$_5$ also contains additional IR active modes, though in our model setup,
these are not polarized along the same direction as the cavity photon mode so do
not couple.

Results for the $\Gamma$ point phonon-polariton spectra of h-BN are presented in
Fig.~\ref{fig:hbn}. Here the resonant $\lambda=0$ IR active modes are two-fold
degenerate due to the three-fold rotational symmetry of the crystal. As we only
incorporated cavity photon modes with a single linear polarization direction the
cavity Hamiltonian breaks this symmetry so that at the resonant frequency one
vibrational mode hybridizes with the cavity mode leading to upper polaritons
(UP) and lower polaritons (LP) while another purely vibrational ``dark'' remains
at the $\lambda=0$ frequency. While the inclusion of electronic response $\chi$ is
observed to lead to an increase in photon character of the lower polariton with
increases in $\lambda$, in h-BN this is seen to occur even when artificially setting
$\chi$ to zero as can be seen from the lighter color curve of the lower polariton
in the bottom left panel of Fig.~\ref{fig:hbn}. Aside from the effective
detuning of the cavity mode due to electronic response this change in mode
character as coupling strength is increased can arise from mixing with other
modes of the system, but also from the change in effective vibrational energy
due to the ionic DSE. This latter effect becomes more prominent (in terms of
percent change of energy level) when the ratio of ionic DSE to the $\lambda=0$
vibrational frequency is larger. This happens for systems with larger Born
effective charges and lower frequency resonant IR vibrational modes.

Results for the $\Gamma$ point phonon-polariton spectra of HfS$_{2}$ are presented in
Fig.~\ref{fig:hfs2}. The underlying physics of this system is similar to that of
h-BN, in that a three-fold rotational symmetry leads to a similar splitting of
the degenerate phonons into UP, LP, and dark modes. However the differences in
$\lambda=0$ spectra and $\chi$ lead to a qualitatively distinct spectra. Compared to the
other systems studied in this work HfS$_{2}$ has relevant IR modes with
frequencies an order of magnitude lower and dielectric response is 3 to 10 times
larger. The lower initial frequency leads to a larger percentage increase in the
upper polariton mode frequency due to the ionic contribution of the DSE, as here
this energy is not as small compared to the energy of the relevant vibration
modes. In the lower-left panel where only a single photon mode is incorporated
this effect is seen to lead to a positive curvature increase in upper polariton
frequency with $\lambda$, however the inclusion of electronic response $\chi$ (right
panels) suppresses this feature by changing the $\lambda$ dependence of the
$\Delta R_{i}\Delta R_{j}$ coefficient from $\approx\lambda^{2}$ to $\lambda^{2}/(1+\chi \lambda^{2})$ as can be seen
in Eq.~\ref{eq:model}.


\section{Conclusion}

In conclusion, we have developed an {\it ab initio} approach for computing the
vibro-polaritonic spectra of cavity-coupled matter systems from the matter
response to nuclei displacements and applied electric fields. This approach
produces results equivalent to previous work implementing a custom Cavity
Born-Oppenheimer approximation (CBOA) energy
functional\cite{flick18_cavit_correl_elect_nuclear_dynam, bonini2022} into DFT.
In contrast, the method presented in this work uses commonly implemented density-functional perturbation theory results to parameterize the
simple harmonic model in Eq.~{\ref{eq:model}} where all cavity parameters appear only in
the model. This is achieved by making use of a mapping between the established
finite electric field enthalpy functional in DFT calculations\cite{Souza2002}
and the CBOA energy functional. We illustrate the method with example
calculations of the vibro-polaritonic spectra of molecular systems CO$_2$ and
Fe(CO)$_5$ and the 2D solids h-BN and HfS$_2$.
While we use DFPT results to parameterize Eq.~{\ref{eq:model}} we note that the
parameters of the matter subsystem (polarizability, Born effective charges, and
vibrational modes) can also be accessed by (non-cavity) experiments so that
experimental results may be compared to those of the model for the interpretation of
polaritonic spectra. Unlike simple two-level models Eq.~{\ref{eq:model}} can
capture asymmetric and nonlinear splitting and other physics. Deviations from
the model may then indicate effects beyond the dipole, mean-field, CBO
approximations or effects of collective coupling.
This relation between electric field response and previous CBOA formulations
presented in this work has implications for the understanding of the physics
captured in the CBOA, practical implementation of CBOA calculations, and the
interpretation of experimental measurements.

\section{Acknowledgements} We acknowledge funding from NSF via grant number EES-2112550 (NSF Phase II CREST Center IDEALS) and startup funding from the City College of New York. All calculations were performed using the computational facilities of the Flatiron Institute. The Flatiron Institute is a division of the Simons Foundation.

\bibliography{cboa_periodic.bib}{}
\setcounter{figure}{0}
\renewcommand{\figurename}{Fig.}
\renewcommand{\thefigure}{S\arabic{figure}}
\begin{figure*}[ht]
  \centering
  \includegraphics[width=\linewidth]{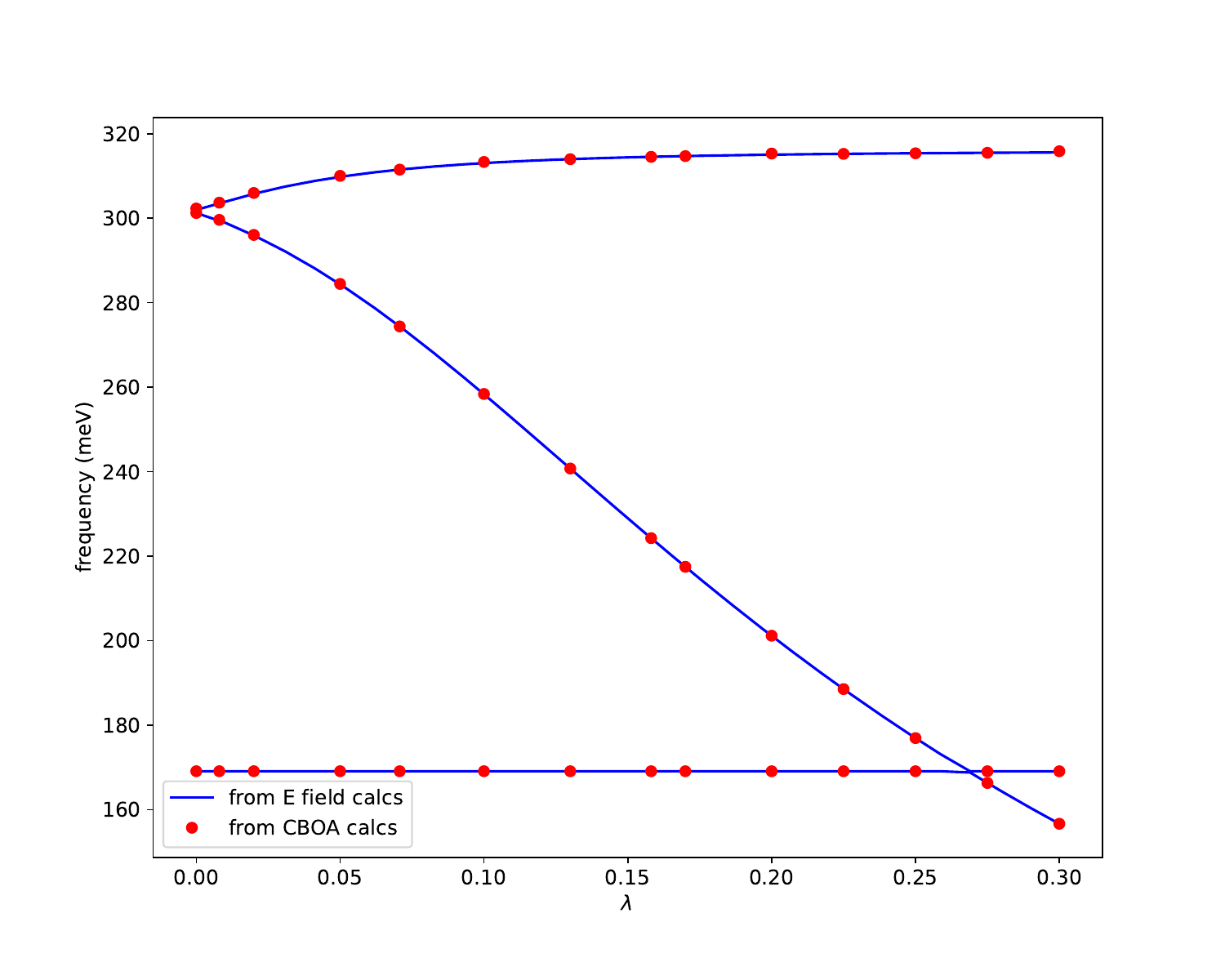}
  \caption{
	Comparison of
  	vibro-polariton splitting in CO$_{2}$ between explicit inclusion of
  	$V_{\mathrm{dipole}}$ in to DFT energy functional (red points) and
  	equivalent results using the model of Eq.~\ref{eq:model} (blue curves).}
\end{figure*}

\end{document}